# Against Spyware Using CAPTCHA in Graphical Password Scheme

Liming Wang, Xiuling Chang, Zhongjie Ren,
Haichang Gao, Xiyang Liu
Software Engineering Institute
Xidian University
Xi'an, Shaanxi 710071, P.R.China
Hchgao@xidian.edu.cn

Uwe Aickelin
School of Computer Science
The University of Nottingham
Nottingham, NG8 1BB, U.K.
uxa@cs.nott.ac.uk

*Abstract*—Text-based password schemes have inherent security and usability problems, leading to the development of graphical password schemes. However, most of these alternate schemes are vulnerable to spyware attacks. We propose a new scheme, using CAPTCHA (Completely Automated Public Turing tests to tell Computers and Humans Apart) that retaining the advantages of graphical password schemes, while simultaneously raising the cost of adversaries by orders of magnitude. Furthermore, some primary experiments are conducted and the results indicate that the usability should be improved in the future work.

## I. INTRODUCTION

A key area in security research and practice is authentication, the determination of whether a user should be allowed to access to a given system or resource. Generally, the most common and convenient authentication method is the traditional alphanumeric password. However, their inherent security and usability problems [6-11] led to the development of graphical passwords as an alternative. To date, there have been several graphical password schemes, such as [7, 18, 20-26]. They have overcome some drawbacks of traditional password schemes, but most of the current graphical password schemes remain vulnerable to spyware attacks.

Commonly, a spyware is a software that, from a user's perspective, covertly gathers information about a computer's use and relays that information back to a third party [1]. Spyware has gradually become one of the most common security threats to computer systems. Password collection by spywares has rapidly increased [4, 5, 12, 13, 15]. The research community has expended much effort [4, 16, 17, 18, 20, 26] on this topic. However, how to protect passwords effectively against spyware attack continues to be a problem. Observing that a practical spyware attack is done by an automated program, we propose a new approach where CAPTCHA is exploited.

CAPTCHA (Completely Automated Public Turing tests to tell Computers and Humans Apart) is a program that generates and grades tests that are human solvable, but beyond the capabilities of current computer programs [27]. The robustness of CAPTCHA is found in its strength in resisting automatic adversarial attacks, automatic adversarial attacks, and it has many applications for practical security, including online polls, free email services, search engine bots, worms and spam, and preventing dictionary attacks [27]. Our proposal creates an innovative use of CAPTCHA in the context of graphical passwords to provide better password protection against spyware attacks.

In this paper, we have proposed a new authentication scheme combining graphical passwords with text-based CAPTCHA. The scheme is easy for humans but makes it almost impossible for automated programs to harvest passwords. The novel scheme is friendly for legitimate users, while simultaneously raising the time and computer capacity cost to adversaries by several orders of magnitude. Experiments showed its effectiveness, but also indicated further research would improve its usability.

The rest of the paper is organized as follows. Section 2 briefly reviews related work. Sections 3 and 4 present our scheme and analyses its security. Section 6 provides the results of experiments described in section 5. Section 7 discusses additional observations and possible extension to our scheme. Conclusions and future work are addressed in section 8.

## II. RELATED WORKS

Most current graphical password schemes, such as [7, 21, 23, 24, 25], require users to enter the password directly, typically by clicking or drawing. Hence, passwords are easily exposed to a third party who has the opportunity to record a successful authentication session. There have been a few graphical password schemes devoted to secure passwords against spyware attacks. In the following, several representatives will be described.

Man, et al [20] proposed that users remember a number of text strings as well as several images as pass-objects. To pass the authentication, users should enter the unique codes corresponding to the displayed pass-object variants and a code indicating the relative location of the pass-objects in reference to a pair of eyes. It is relatively hard to crack this kind of password, but the complex memory requirement is an obstacle to its popularity.

In [26], users need to recognize pass-objects and click inside the convex hull formed by all of the pass-objects. If properly designed, this method can provide good security. However, from time to time the convex hull is either too small to click or too large, creating a guessing problem. Moreover, to provide a large password space may result in a crowed screen and indistinguishable objects. The method in [22] to resist shoulder-surfing is a trivial trick, where a user must click a group composed of both the pass-object and decoy-object rather than click the pass-objects directly. The prototype presented in [22] does not provide sufficient security, having only two objects in each group.

In 2006, Weinshall proposed another challenge-response protocol that relied on a shared secret set of pictures [18]. To reduce the amount of information given out with each authentication session, the image set memberships are used to select a certain path on an image mosaic, with the user providing only a code that depends on the path's endpoint. This scheme was claimed to be so strong that an observer who fully records any feasible series of successful interactions could not compute the user's password. However, it was demonstrated by Golle and Wagner [19] that the attacker can learn a user's secret key with a SAT solver after observing as few as six successful user logins.

In essence, the above methods adopt a challenge-response protocol to confuse the spyware. They can prevent the passwords being cracked by the spyware and falling into the hand of an adversary, along with resisting replay attacks. Taking the previous mechanisms for reference, our scheme also uses a challenge-response protocol to enhance security. But, unlike these methods, our scheme innovatively applies CAPTCHA to graphical passwords to create a highly secure authentication method.

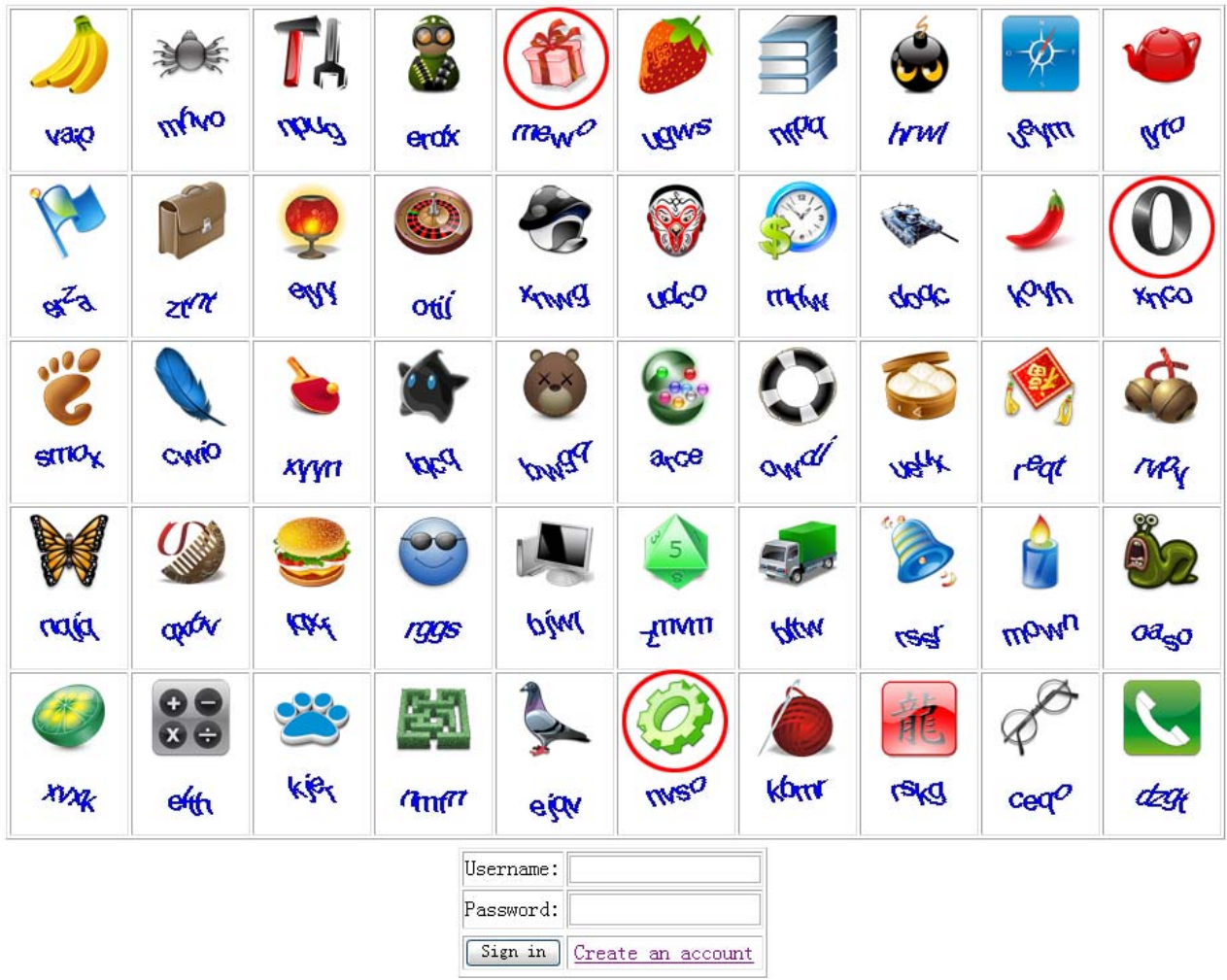

Figure 1. The interface of the basic scheme (The pass-images are circled).

### III. OUR SCHEME

Our approach is motivated by the observation that effective spyware attacks are launched from automated programs. We realized that to increase security passwords should be accompanied by a product of a "computation" that is difficult for machines. As an authentication method, the scheme should also be user friendly. Considering these requirements, we applied CAPTCHA to graphical password schemes.

CAPTCHA is a program designed to test whether the user is a computer or a human, by creating a task easy for humans but difficult for machines [27]. It is based on hard AI problems which cannot be solved with any greater accuracy than what is currently known to the AI community [31]. CAPTCHA is now almost a standard security mechanism for addressing undesirable or malicious Internet bot programs [28] and major web sites such as Google, Yahoo and Microsoft all have their own CAPTCHAs. The state-of-the-art CAPTCHAs mainly include three types: text-based schemes, sound-based schemes and image-based schemes. The most widely deployed schemes are text-based CAPTCHAs and we also use this in our schemes.

After introducing a basic scheme with a hidden safety loophole, we will describe an improved scheme that is designed to fill the hole. The performances of the both schemes depend extremely on the property of CAPTCHA.

*A. The Basic Scheme*

The basic scheme embeds a text-based CAPTCHA into a simple graphical password scheme. Each image has a CAPTCHA instance called adjunctive string and the strings are generated at random by the system. In the register phase, users are required to select and remember images as their password images (pass-images). To be authenticated, users need to distinguish his/her pass-images as well as solve a test by recognizing and typing the adjunctive string below each pass-image. For example, in Figure 1, assume the three images with red circles are pass-images, users should input the adjunctive strings 'mewo', 'xnco' and 'nvso' correctly to pass the authentication.

For simplicity, we assume that the CAPTCHA here is an ideal CAPTCHA that is hard enough for machines to recognize while easy for humans to solve.

In the case that adversaries are automated programs without human intervention, the scheme has a strong resistance to replay attack. Namely, even if it observes a successful login, a spyware program cannot launch a replay attack. This can be illustrated from two aspects. Firstly, pass-images are entered by typing random adjunctive strings rather than clicking directly. In other words, the entered strings are the trap instead of the real password. Secondly, machines have no ability to recognize the characters embedded in each image. It follows that it is rather difficult for an automated program to find pass-images according to the recorded strings.

The loophole in this scheme occurs if the adversary is a person and the spyware is an assistant. The password will be in danger because CAPTCHA is easy for a person. In this case, the person can see what the spyware has gathered, a successful login scene along with the entered characters. Then, a person can crack the passwords without much effort. For 26 lower case letters in the scheme, the probability that different images have the same string is 1/456976, which can be ignored. One useful method for password cracking is to divide the gathered strings with four characters into groups and then compare each segment with that below each image. To close this loophole, we constructed an improved version.

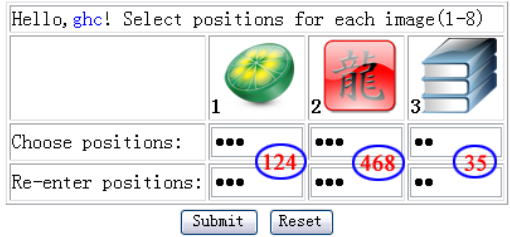

(a) The interface of register.

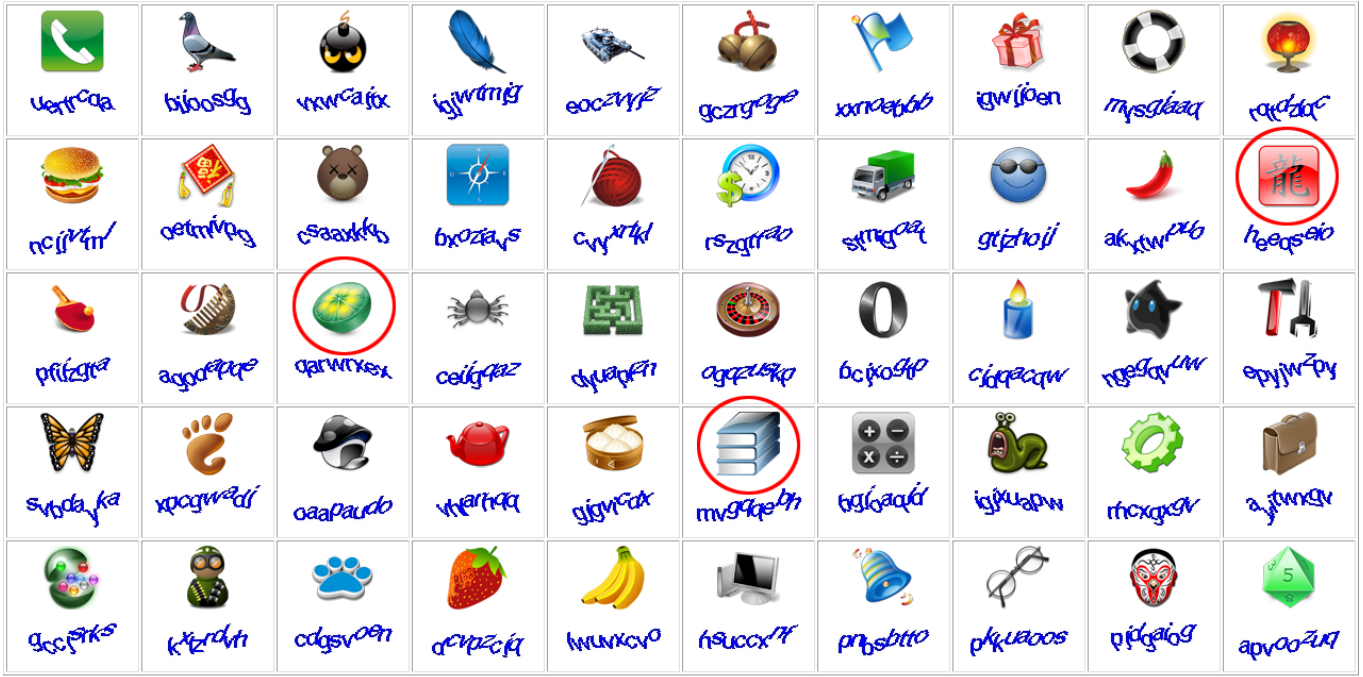

(b) A login screen of the same user 'ghc' in (a).

Figure 2. The interface of the basic scheme (The pass-images are circled).

## B. The Imporved Scheme

The vulnerability of the basic scheme lies in two factors. One is the requirement that CAPTCHAs should be human user friendly. The other is the reversible relationship between passwords and what is entered. That is, pass-images determine what is entered and vice versa. What's more, we noted that the reversible relationship depends greatly on the fact that the probability of different images with the same adjunctive string is close to zero and that the trap of each pass-image has a uniform length. While the former is necessary for a popular authentication scheme, we are encouraged to disturb the latter.

One possible method is increasing the probability by decreasing the types of letters or the length of adjunctive string. This method might work, but it will increase the probability of illegal login by random guessing. Thereby, it is ineffective as a security method. Our alternative is to replace the uniform length with a random one predefined by users. In other words, the number of characters entered is determined by users.

In our improved scheme, users are required to select and remember letter positions, ie choose several specific letter positions within a string of letters; for example, letters in $1^{st}$, $4^{th}$ and $5^{th}$ position in the string will become the code. These letter positions are the called pass-positions for each pass-image. During the authentication, users should enter the characters shown in the pass-positions of each pass-image. An example is shown in Figure 2.

In Figure 2(b), the three circled images are pass-images, the strings with them are 'qarwrxex', 'heeqseio', and 'mvgqqebh' respectively, and the corresponding pass-positions are (1, 2, 4), (4, 6, 8), and (3, 5) shown in Figure 2 (a). A user can input any combination of the three sequences, 'qaw', 'qeo', and 'gq' to be authenticated successfully.

This scheme is strongly resistant to attacks launched by humans with spyware, while simultaneously preserving the advantages of graphical password schemes. The related security analysis will be given in the following section and usability problems will be discussed in Section 5, 6 and 7 through experiments.

## IV. SECURITY ANALYSIS OF THE IMPROVED SCHEME

### A. Capability to Withstand Spyware

There are many different kinds of spyware [1, 2], such as browser hijackers, keyloggers and spybots. We have focused on the spyware cluster that runs in the background collecting passwords. The security of our scheme relies on the robustness of CAPTCHA in resisting automatic adversarial attacks. However, it is not clear whether there is a true CAPTCHA at all and some reports show that some text-based CAPTCHAs can be partly or almost broken by automatic programs [3, 29, 30]. With the assumption that spyware is capable of detecting and recording screen snapshots, entered strings and the system feedback, we will analyze the security of the improved scheme from two extreme aspects. Firstly, it is impossible for machines to solve the CAPTCHAs in our scheme, the ideal case. Secondly, CAPTCHAs can be completely solved by machines, the worst case.

Under ideal conditions, spywares have no chance of gaining the passwords without human invention, similar to the discussion in sections 3.1. If people are involved, spyware assistance can help users to break the scheme. What the spyware needs to do is to catch the password string entered by the legal user. To crack passwords, adversaries should solve the CAPTCHA himself or by employing human workers. It is costly to obtain a password because the pass-positions of each pass-image are unknown and thereby it is hard to manually find the correspondence between pass-images and what is entered. Even for the lowest level security, adversaries must recognize 400 CAPTCHAs. In this case, there are three pass-images, each with a pass-position and then the attacker can easily divide the entered string into three segments each with a specific character. The probability of a letter displayed below one image is $1-\left(\frac{25}{26}\right)^8 \approx 0.27$. For each authentication, there are 100 images on screen in our scheme with about 27 images which have a common specific character. That is, there are 27 candidates including a pass-image and 26 decoys. This illustrates that the attacker can gain a pass-image with a probability of $\dfrac{1}{100\times\left(1-\left(\frac{25}{26}\right)^8\right)} \approx 0.037$ and can penetrate the passwords with a probability of $\left(\dfrac{1}{100\times\left(1-\left(\frac{25}{26}\right)^8\right)}\right)^3 \approx 0.0000512$ approximately from one observation and analysis. Through interaction, the attacker can gradually get rid of all the decoys. For the second observation, the number of decoys will be $26\times\left(1-\left(\frac{25}{26}\right)^8\right) \approx 7$. After the third observation, there will only be about three CAPTCHAs which contain the specific character. The attacker can find the users passwords correctly in four sessions. So the attacker must solve approximately 400 CAPTCHAs and conduct many observations and comparisons, which is time consuming and costly. More complex work is required if the correspondence between pass-images and entered strings are unknown. Therefore, our scheme has a strong resistance against spywares under the ideal environment.

Projecting the worst condition, that CAPTCHAs can be completely solved by machines, it is possible that spywares could crack passwords because each successful login reveals some information about the password. One method is to divide the entered strings into different segments and find the passwords from images which contain the same segments from analyzing different login sessions. Another method is to find the common images by excluding images without any character of the entered string. For instance, when the passwords lie in the lowest security level, it is possible to crack the passwords in four sessions, as discussed above.

This worst case scenario is not probable, unless spywares can gather sufficient information in the background and can break CAPTCHAs quickly. Currently, no programs can break a CAPTCHA automatically in a short time. Furthermore, even if the currently applied CAPTCHAs are effectively broken, there will always be versions with higher security in production. In addition, as long as the hard AI problems underlying CAPTCHA are unsolved, successful attacks will advance the development of more robust CAPTCHAs.

Therefore, it is demonstrated that our scheme is secure against spyware as long as CAPTCHAs can not be broken by automated programs. Any defeated CAPTCHAs will be substituted by more robust ones. If humans are involved, the cost of cracking a password is significantly increased.

*B. The Size of the Password Space*

Now, we consider the raw size of the password space, assuming users are equally likely to pick any element as their password. According to the definition in [23], the raw size is an upper bound on the information content of the distribution that users choose in practice.

We compute the size $S(L, N, M)$ of password space of total entered length equal to $L$ when there are $N$ images displayed and the length of CAPTCHAs is equal to $M$. In our scheme, for security reasons, the number of pass-images is required to be not less than 3. Thus, $S$ is defined in terms of $P(K, L, N, M)$, the number of passwords with number of pass-images equal to $K$ by:

$$S(L, N, M) = \sum_{K=3}^{L} P(K, L, N, M) \quad (1)$$

In turn, $P(K, L, N, M)$ can be defined in terms of $O(K, L, N, M)$, the number of passwords when the K pass-images have been confirmed, by:

$$P(K, L, N, M) = C_N^K \cdot O(K, L, N, M) \quad (2)$$

The reason is that the $K$ pass-images have no relative order. Assume the number of pass-positions for one pass-image is $n$, we can get,

$$n_1 + n_2 + \cdots + n_K = L \quad (3)$$

Here, the problem can be seen as an issue of the ordered partitions of positive integer. $L$ is partitioned into $K$ ($1 \leq K \leq L$) sections. According to the theorem of the partition of positive integer, the generating function of sequence of partition numbers is $(\sum_{j=1}^{M} x^j)^K$. We assume that there are $G(M, K, L)$ different partition situations in all, and any one partition can be denoted by:

$$F_i : n_{1i} + n_{2i} + \cdots + n_{Ki} = L \quad (i = 1, 2, \ldots, G) \quad (4)$$

Then, $O(K, L, N)$ can be defined in terms of n by:

$$O(K, L, N, M) = \sum_{i=1}^{G} \left( \prod_{q=1}^{K} C_M^{n_{qi}} \right) \quad (5)$$

Combining the formulae, we can compute the size of the password space. The results for the password space are given in Table 1, when $N$=50, $M$=8, and $3 \leq L \leq 10$.

Table 1 results are encouraging. however, that is the raw size of our password space. In practice, actual password space will be reduced due to users' individual preferences. Additionally, the size of the password space of our scheme is actually smaller than that of text-based passwords (94 printable characters available) when the length is equal to or greater than 10 ($94^{10} \approx 5.4 \times 10^{19}$). As we know, the exhaustive-search attack is always produced automatically by software rather than by people. In our scheme, CAPTCHA is introduced to resist this kind of attack. Subsequent CAPTCHA development maintains the security of our method, as each round of development becomes more difficult for automated cracking programs and more expensive for manual, human-based cracking programs.

TABLE I. NUMBER OF PASSWORDS OF ENTERED LENGTH EQUAL TO L ($N$=50 AND $M$=8).

| L | Password space size | log$_2$(#space size) |
|---|---|---|
| 3 | $1.0 \times 10^7$ | 23.3 |
| 4 | $1.0 \times 10^9$ | 30.0 |
| 5 | $8.3 \times 10^{10}$ | 36.3 |
| 6 | $5.5 \times 10^{12}$ | 42.3 |
| 7 | $3.1 \times 10^{14}$ | 48.1 |
| 8 | $1.5 \times 10^{16}$ | 53.8 |
| 9 | $6.6 \times 10^{17}$ | 59.2 |
| 10 | $2.6 \times 10^{19}$ | 64.5 |

*C. Brute Force Attacks*

Brute force attack, trying to randomly guess the correct passwords, is the simplest form of attack for an authentication scheme. For our scheme, with a candidate set of A characters, the probability that a single random guess succeeds is $K!/A^L$.

For one legitimate user, every time to authenticate, there are $K!$ choices of entered string, since pass-images have no relative order. Just as the instance shown in Session 3.2, the user can enter any combination of three sequences to authenticate. Thus, there are six possible strings to enter, 'qawqeogq', 'qawgqqeo', 'qeoqawgq', 'qeogqqaw', 'gqqawqeo', 'gqqeoqaw'. For $(A, L, K) = (26, 8, 4)$, we obtain $4!/26^8 \approx 1/26^7$. The attacker has a very low probability of logging on successfully with a brute force attack.

V. EXPERIMENTAL MENTHODOLOGY

During the testing phase, fifty images of 60×60 pixels and corresponding CAPTCHAs were displayed on the screen in the prototype of the improved scheme. All the images were downloaded from http://www.chinaz.com freeware website and processed for study only. The length of CAPTCHA strings was 8, and the characters contained 26 lowercase letters. The CAPTCHA algorithm was designed to generate crowded, distorted and rugged strings similar to the CAPTCHA being used in Google email service for its acknowledged robustness.

A total of 36 participants were invited to complete the experiments and answer some questions. The participants, of whom there were 15 women and 21 men, were staff and students from a university community and unfamiliar with our scheme. The average age of the participants was 27 years (StdDev=4.5), and ranged from 21 to 39 years. All the participants were required to complete the following operations individually.

Firstly, they need answer a demographic questionnaire, which collected information including age, sex, highest degree earned and computer experience. At this session the scheme and procedures for the experiments were explained to them in detail.

Secondly, the user was required to select three or more pass-images. After selecting the pass-images, the user set the pass-positions for each image. During the testing phase, if the participants forgot the pass-images or the pass-positions, the password which they have just set was shown to them.

In the testing phase, the data were collected longitudinally: first, at end of the training session (P1), then one week later (P2), and finally one month later (P3). For P1, each participant was asked to set a password, and authenticate ten times. For P2 and P3, if a participant entered an incorrect password, he or she was allowed to re-enter the password. Three login attempts were permitted for each participant.

## VI. RESULTS

### A. The Mean Success Login Percentage

In P1 testing session, 9 of 36 participants completed with no mistakes in ten times of login, while the others, to a greater or less extent, made some incorrect submissions. The mean success login percentage is 87.8% (StdDev=9.29). The reasons offered by the participants for the incorrect submissions included difficulty in identifying the text-based CAPTCHAs generated by our algorithms and sometimes in locating the exact pass-positions.

### B. The Mean Login Time

In P1 testing session, the mean login time of all participants is 22.04 seconds (StdDev=10.9) which is acceptable for most participants. The results show that there is a significant difference in terms of time to respond to a challenge ($F(35,280) = 15.48$, $p<0.01$). The main reason may be that the CAPTCHAs are randomly generated so that sometimes they are easy to recognize but sometimes more difficult. As the images are randomly located, the time for recognition also differs. Results show that the majority of participants chose three to five pass-images, with only three participants choosing more than five pass-images. Mean times and standard deviations of logins with different pass-images are shown in Table 2.

### C. Password Memorability

In P2 testing session, 80.6 percent of participants successfully logged into his/her account in three attempts, and in P3 session, 72.2 percent participants were successful. Interviews with participants provided the following reasons for memory lapses: a) the difficulty of remembering the pass-positions and b) the difficulty of remembering the relationships between pass-positions and pass-images.

TABLE II. MEAN TIMES (SECONDS) AND STANDARD DEVIATIONS OF CHALLENGES WITH DIFFERENT PASS-IMAGES

| Numbers of pass-images | Numbers of persons | Mean | StdDev |
|---|---|---|---|
| 3 | 18 | 17.57 | 7.7 |
| 4 | 12 | 24.76 | 6.7 |
| 5 | 3 | 44.00 | 16.8 |
| 6 | 1 | 25.87 | 4.5 |
| 7 | 1 | 16.87 | 3.6 |
| 9 | 1 | 15.37 | 5.4 |

## VII. DISCUSSION

In comparison to other graphical password schemes, such as [7,14,26], there are some advantages and disadvantages in our improved scheme. One disadvantage is that it is more complex and increases users' memory load. Users have to remember both the pass-images and pass-positions. To be authenticated, users need to recognize the pass-images and input the characters of the text-based CAPTCHAs on the pass-positions correctly. These factors have increased the complexity of the login process. However, although it is complex and cumbersome, the improved scheme is strongly resistant to spywares, which is our primary focus.

A comparison of login time for our scheme shows that, our scheme, as other graphical schemes, is longer than that of text-based schemes. However, when compared to other graphical password schemes our login time is shorter. For instance, the mean login time of CHC is 72 seconds and Déjà vu is 27 to 32 seconds because there are multiple rounds of challenges in these schemes [26]. In [18], a typical entry takes over 3 minutes using a high-complexity protocol and over 1.5 minutes with a low-complexity protocol. Moreover, schemes against spywares [18, 20] also challenge user's memory capacity to a great extent. In [18], the high-complexity protocol asks the user to remember 30 pictures. And in [20], the user needs to remember 16 random strings for corresponding 16 pass-images. The mean login time of our improved scheme is 22.04 seconds. We believe that our login times will decrease with familiarity with the scheme. All experiments were undertaken in lab and all the participants were new to our scheme. The users' login speed should be faster with the extended use.

If the scheme is moved to real usage, the settings of the parameters can be adjusted to adapt to different security demands and application situations. There are $M$ images randomly generated including $N$ pass-images, and there are $S$ rounds of challenges for one login. For each round, $m$ images are displayed with $n$ pass-images. With the increasing of the numbers of total images, pass-images and length of text-based CATPCHAs, the security of the scheme will be enhanced. For example, when $M = 250$ and $N = 10$, the spyware will detect 10 pass-images and the corresponding pass-positions for 250 images. This requires recording of hundreds of logins and recognition of a huge number of CAPTHCAs. Gathering so much information may take a long time and recognizing the CAPTCHAs also needs an extensive manpower. Certainly, increasing the setting for high security is at the expense of usability.

There are also some user behaviors which create risks for our scheme. First, the passwords selected by user often accord with a particular trend. For example, in order to make the password easily remembered, most users select the same position for different pass-images, first or anterior positions, consecutive positions or one position for each pass-image. And certain images were selected by a number of users as pass-images. All the factors mentioned above can reduce the practical password space and increase the possibility of "guessing" attacks.

Second, we find that there is always a significant time gap when entering characters belonging to two different pass-images. The reason is that users are used to enter corresponding characters after he finds a pass-image. Such a situation will be recorded and utilized by spywares. This problem can be solved by entering characters by turns which belong to different CATPCHAs in a certain order.

In summary, our improved scheme is resistant to spyware attack, and the rules for setting passwords have increased the cost and time of the human intervention attack.

## VIII. Conclusion and Future Works

In this paper, we have presented a new approach to protect user's password against spyware attack. Our main contribution is that we introduce CAPTCHA into the realm of graphical passwords to resist spyware programs. From a security viewpoint, this exploration is expected to advance the development of graphical passwords. While the design of CAPTCHA is an interdisciplinary topic and the current collective understanding of this topic is still in its infancy, we do not claim that our scheme is immediately feasible. However, we believe that our method will enhance current security and as CAPTCHA increases in effectiveness our method will also increase computer security.

The results of our experiments show that the future research should concentrate on improving the login time and memorability. Furthermore, when a user inputs the corresponding substrings which belong to different CAPTCHAs, the time gap is longer than the time between two characters in one substring. So a method for narrowing the time gap in the entering process and reduction of the impact of user's choice trend on security, provide other areas for future research.

## Acknowledgment

The authors would like to thank the reviewers for their careful reading of this paper and for their helpful and constructive comments. Project supported by National Natural Science Foundation of China (60903198) and Natural Science Basic Research Plan in Shaanxi Province of China (SJ08F25).